\documentclass[twocolumn]{aastex7}
\usepackage{amsmath,amssymb,bm}
\usepackage{booktabs,graphicx,array,tabularx,multirow,url}
\newif\iftracked
\trackedtrue
\newcommand{\rev}[1]{\iftracked{#1}\else#1\fi}
\newcommand{\mbh}{M_\bullet}
\newcommand{\Msun}{M_\odot}
\newcommand{\dex}{\,\mathrm{dex}}
\newcommand{\normal}{\mathcal N}
\newcommand{\trans}{\mathsf T}
\newcommand{\expt}{\operatorname{E}}
\newcommand{\Rinfo}{R_{\rm info}}
\newcommand{\Istep}{I_{\rm step}}
\newcommand{\sigop}{\sigma_{R,\rm op}}

\newcommand{\CalBXMed}{0.71}
\newcommand{\CalBXPlus}{0.13}
\newcommand{\CalBXMinus}{0.14}
\newcommand{\CalBMMed}{0.74}
\newcommand{\CalBMPlus}{0.21}
\newcommand{\CalBMMinus}{0.23}
\newcommand{\CalWidthMed}{1.07}
\newcommand{\CalWidthPlus}{0.20}
\newcommand{\CalWidthMinus}{0.16}
\newcommand{\OperatingWidth}{1.11}
\newcommand{\HistRawMed}{0.68}
\newcommand{\HistRawLo}{0.17}
\newcommand{\HistRawHi}{1.15}
\newcommand{\HistInvariantMed}{0.71}

\newcommand{\SeqReachThirtyTwo}{66.2}
\newcommand{\SeqOverallThirtyTwo}{46.3}
\newcommand{\SeqReachFortyEight}{77.5}
\newcommand{\SeqOverallFortyEight}{50.3}

\graphicspath{{./}}
\hypersetup{hypertexnames=false,
  pdftitle={The Unresolved High-Mass Frontier of the Black Hole Fundamental Plane},
  pdfauthor={Ramananda Santra}}
\shorttitle{A Direct-Mass Test of the High-Mass Fundamental Plane}
\shortauthors{Santra et al.}

\begin{document}


\title{The unresolved high-mass frontier of the black hole Fundamental Plane}

\author[0009-0002-0373-570X]{Ramananda Santra}
\email{ramananda.santra@icts.res.in}
\affiliation{International Centre for Theoretical Sciences, Tata Institute of Fundamental Research, Bengaluru, Karnataka 560089, India}

\author[0000-0002-5730-0376]{Samik Mitra}
\email{samik.mitra@icts.res.in}
\affiliation{International Centre for Theoretical Sciences, Tata Institute of Fundamental Research, Bengaluru, Karnataka 560089, India}

\author[0000-0003-0428-2140]{Shahram Abbassi}
\email[show]{sabbassi@uwo.ca}
\correspondingauthor{Shahram Abbassi}
\affiliation{Department of Physics and Astronomy, Western University, 1151 Richmond Street, London, Ontario N6A 3K7, Canada}

\begin{abstract}
\rev{Accreting black holes follow a Fundamental Plane (FP) linking radio and X-ray luminosities to black hole mass, but whether this relation changes in the ultramassive regime, above $\sim10^{10}\,\Msun$, remains unresolved. Brightest cluster galaxies (BCGs) provide the most direct access to this regime, although earlier claims of enhanced radio emission relied largely on heterogeneous nuclear measurements and black hole masses inferred from host-galaxy proxies. We recalibrate the forward FP using galaxies with direct dynamical mass measurements and reassess the evidence for a high-mass radio excess using the available BCG data. We find that the archival proxy-mass sample cannot uniquely establish such an excess because the inferred signal remains sensitive to the mass calibration and radio-selection properties of the sample. Using mock observing programs, we further show that the predictive width measured from existing observations makes a high-mass transition difficult to both detect and localize with present samples. The existing data therefore provide no compelling evidence for a distinct high-mass branch of the FP. Resolving this question will require a substantially larger, uniformly selected sample of galaxies with direct black hole masses and matched, preferably quasi-simultaneous, nuclear radio and X-ray measurements. A focused pilot survey of several dozen direct-mass systems would provide a practical first step toward establishing whether the required precision is achievable.}
\end{abstract}


\section{Introduction}\label{sec:intro}

Accreting black holes spanning nine decades in mass obey a common scaling between compact radio luminosity, hard X-ray luminosity, and black hole mass, known as the Fundamental Plane (FP) \citep{Merloni2003,Falcke2004,Gultekin2009,Gultekin2019}. The physical content of this FP is appealing: if the jet is launched by a scale-invariant magnetohydrodynamic process and the X-rays trace a radiatively inefficient accretion flow, the radio and X-ray outputs should scale together with mass in a calculable way \citep{HeinzSunyaev2003,YuanCui2005}. The FP has accordingly become both a standard tool for low-luminosity accretion and, more ambitiously, a black hole mass estimator, including in searches for intermediate-mass black holes \citep{Karimi2024,Abbassi2026}. \rev{For an inverse FP estimate to have a clear physical meaning, the radio measurement must isolate the compact jet core and the X-rays must arise from the corresponding sub-Eddington accretion flow; a low Eddington ratio alone does not guarantee that these conditions are satisfied \citep{Abbassi2026}.} However, these applications become increasingly fragile at the extreme high-mass end ($>10^{10}\,\Msun$). The fitted slopes and residual scatter depend on source class, accretion state, observing frequency, angular scale, and the time separation of the radio and X-ray data \citep{Kording2006,Plotkin2012}. More importantly, the FP is calibrated as a forward relation: it predicts the radio luminosity expected for a given X-ray luminosity and black hole mass, and therefore cannot simply be inverted to yield a mass. Such an inversion additionally depends on the range of X-ray luminosities and masses represented in the sample and on how the objects entered the sample \citep{Kelly2007,FeigelsonBabu2012,Gelman2013}.

Brightest cluster galaxies (BCGs) are where these issues bite hardest and where the astrophysical stakes are highest. BCGs host the most massive black holes known \citep{McConnell2011,McConnell2012,Thomas2016,Mehrgan2019}, sit at the bottom of deep cluster potentials, and are almost always radio sources whose compact cores are embedded in jets, lobes, and buoyant cavities inflated over many outburst cycles \citep{McNamaraNulsen2007,Hogan2015a,Hogan2015b,Hogan2021}. \rev{Their nuclei accrete at very low Eddington ratios, but low accretion rate alone does not guarantee a unique radio--X-ray coupling: changes in the dominant X-ray-emission mechanism or in compact-core production can move a source away from the standard FP behaviour \citep{YuanCui2005,Abbassi2026}.} A genuine FP break at high mass would therefore have substantive physical implications, indicating a qualitative change in the jet--accretion coupling in the most massive black holes, potentially attributable to high spin, a magnetically arrested disc (MAD) state, or a distinct fueling mode governed by the ambient hot medium \citep{Tchekhovskoy2011}. Whether such a break actually exists, however, has remained an open question.

The existing evidence points toward a discrepancy but falls short of establishing one. Several studies found BCG radio luminosities in excess of the FP prediction and argued that masses above $10^{10}\,\Msun$ would reduce the discrepancy \citep{Hlavacek2012,Mezcua2018}. These results provided the strongest available motivation for investigating the high-mass regime, but they did not constitute a measurement of a mass-localized break. The radio data were assembled from heterogeneous targeted and survey observations. Hard X-ray non-detections were common, and the centimetre beams admitted emission from well beyond the nucleus. Moreover, the masses were derived from a $K$-band scaling proxy. Such proxies are known to be systematically uncertain at the top of the mass function, where the scaling relations themselves are poorly constrained and sensitive to whether luminosity or velocity dispersion is used to calibrate the mass relation \citep{Lauer2007,KormendyHo2013,McConnellMa2013,Kluge2020}. Crucially, an error in the proxy zero point shifts the predicted radio luminosity by almost exactly as much as a genuine population offset does. \rev{The earlier BCG analyses therefore mix two effects that the proxy-based data alone cannot separate: a systematic error in the black-hole mass scale and a genuine radio excess. This observational degeneracy, by itself, is not evidence for a mass-localized break.}

Triaxial Schwarzschild modelling has changed this situation. Sixteen BCGs now have direct dynamical mass constraints, eight centred above $10^{10}\,\Msun$ \citep{deNicola2025,deNicola2026}. The mass axis is therefore measured independently of the relation being tested. This does not, however, provide a measured BCG Fundamental Plane. These galaxies were selected for dynamical modelling, not for homogeneous nuclear radio and X-ray observations, and their existing coverage shares no common physical aperture, rest-frame frequency, or observing epoch.

In this paper, we study what would be required to establish, or rule out, a change in the FP above $10^{10}\,\Msun$. The limiting factors are the predictive width of matched nuclear measurements, the geometry of the target sample, and the availability of a sufficiently large direct-mass sample. Section~\ref{sec:data} describes the three datasets and the distinct role each can play. \rev{Section~\ref{sec:framework} describes how we test whether a high-mass radio departure can be separated from ordinary FP trends and other astrophysical effects.} Section~\ref{sec:results} presents the results. Section~\ref{sec:discussion} discusses the physical implications and the observing strategy they imply, and Section~\ref{sec:conclusions} summarises the findings. Throughout, we write $R=\log_{10}(L_R/{\rm erg\,s^{-1}})$ for the rest-frame 5\,GHz luminosity $\nu L_\nu$, $X=\log_{10}(L_{2-10\,\rm keV}/{\rm erg\,s^{-1}})$, and $\mu=\log_{10}(\mbh/\Msun)$, where $\mbh$ is the mass of the central black hole.

\section{Data}\label{sec:data}
No single existing dataset jointly supplies a homogeneous nuclear radio and X-ray analysis, an unbiased population offset, and a secure high-mass calibration. We therefore treat three datasets as distinct, non-interchangeable inputs to the analysis, each addressing a different part of the problem (Table~\ref{tab:samples}).

\subsection{Direct-Mass Calibration Sample}\label{sec:cal}
The calibration comprises 31 radio--X-ray epochs of 29 galaxies with direct measurements of stellar-, gas-, or maser-dynamical black hole masses, reconstructed from \citet{Gultekin2019}. Its role is to establish how well the centimetre/X-ray FP predicts radio luminosity in practice. It is heterogeneous in beam size, observing frequency, spectral conversion, obscuration and cadence. Such heterogeneity is intentional, as it defines the experiment whose precision can be measured from existing data. \rev{We propagate these effects rather than averaging them away. Distance errors move mass and luminosity together and are treated coherently. Two objects with unusual distance or mass determinations are handled separately: the Galactic Centre, whose mass and distance come from a joint orbital solution, and NGC~4258, which has a geometric maser distance.} This sample establishes the predictive width of the forward relation; it does not, by itself, constrain whether BCGs as a class depart from that relation, which requires the datasets described in the next sections.

\subsection{Proxy-Mass BCG Archive}\label{sec:hist}

The archive contains 72 BCGs with 42 hard X-ray detections and 30 upper limits \citep{Mezcua2018}, each with a 5-GHz-equivalent luminosity and flux density, a $2$--$10$\,keV constraint, a redshift, and a $K$-band mass proxy; five transcription errors were corrected against the source table.\footnote{Corrected hard-band values: RXJ0058.9$+$2657 (41.06), RXJ0341.3$+$1524 (40.84), RXJ1304.3$-$3031 (39.86, upper limit), RXJ1604.9$+$2356 (40.45), and RXJ0352.9$+$1941 (photometric, 42.90, from a transcribed 52.9).} It has been used to establish what a selected, proxy-based sample can say about a BCG radio offset. \rev{The X-ray upper limits are included directly in the likelihood, so non-detections contribute information without being treated as detections.} We do not attempt a completeness correction: the parent list needed to reconstruct the inclusion function does not exist, since objects entered through a mixture of targeted cluster programs and archival availability. Separating a genuine radio offset from proxy-mass error requires masses obtained independently of the $K$-band scale, which motivates the direct-mass BCG sample described next.

\subsection{Direct-Mass BCG Sample}\label{sec:bcgs}

This sample consists of 16 galaxies from \citet{deNicola2025,deNicola2026}, and defines the mass range of the \rev{future observing test}. Because object-level dynamical likelihoods are not yet available, \rev{we approximate each published asymmetric mass interval by a two-sided Gaussian distribution in $\log\mbh$, with a $0.10\dex$ systematic floor} (Table~\ref{tab:bcgs}); these are working approximations, not reconstructions of the orbit-model posteriors, and should be replaced when the likelihoods are released. Two features matter for everything that follows: all 16 are BCGs, so a cluster-environment indicator is constant and cannot be separated from the overall zero point; and no homogeneous accretion-state classification exists for them. The forecast conditions on these published masses and does not model the selection function for obtaining a dynamical mass.

\begin{table}[!ht]
\centering
\footnotesize
\caption{The three datasets and their primary role in the analysis.\label{tab:samples}}
\begin{tabular}{lcl}
\hline
Sample & Size & Primary role \\
\hline
Direct-mass hosts & 29 & FP forward precision \\
Proxy-mass BCGs & 72 & Proxy radio offset \\
Direct-mass BCGs & 16 & Future mass coordinate \\
\hline
\end{tabular}
\tablecomments{The calibration sample does not measure a BCG FP; the archive does not measure an unbiased population offset; and the 16 BCGs do not yet have a homogeneous nuclear radio/X-ray relation.}
\end{table}

\section{High-Mass Fundamental Plane Tests}\label{sec:framework}

\rev{The three datasets in Section~\ref{sec:data} answer different astrophysical questions and are therefore kept separate. The direct-mass calibration sample tells us how accurately the observed centimetre/X-ray FP predicts the radio luminosity of an individual nucleus. The 72-BCG archive tests whether the previously reported BCG radio excess can be separated from uncertainty in the proxy mass scale and from radio selection. The 16 direct-mass BCGs define the mass range over which a future experiment could search for a genuine change in the FP. Our analysis is organized around these three questions. The analysis propagates the relevant measurement uncertainties and translates a proposed observing program into the probability of detecting a high-mass transition. Mathematical details that are not needed for the astrophysical interpretation are given in Appendix~\ref{app:methods}.}

\subsection{Forward FP Calibration}\label{sec:bayes}

We first establish how precisely the centimetre/X-ray relation predicts radio luminosity from the 29-host calibration sample. We model the FP as a linear relation in $X$ and $\mu$ about a fiducial pivot,
\begin{equation}
R=38+a+b_X(X-40)+b_M(\mu-8)+\varepsilon,
\label{eq:fp}
\end{equation}
\rev{where $a$ is the zero point and $b_X$ and $b_M$ are the X-ray and black-hole-mass slopes. Both $X$ and $\mu$ are measured with finite uncertainties, so we do not treat their tabulated values as exact. Instead, the fit allows the underlying X-ray luminosity and black hole mass of each object to vary within its measurement constraints while solving simultaneously for the FP coefficients, following the measurement-error treatment commonly used for astronomical scaling relations \citep{Kelly2007,FeigelsonBabu2012}. Asymmetric errors and the common distance dependence of mass and luminosity are retained, and repeated observations of the same host are not treated as independent galaxies. The quantity needed for the high-mass experiment is the total predictive width $\sigma_R$: the expected dispersion in radio luminosity when the fitted FP is used to predict a new nucleus. The present 31 epochs do not robustly separate source-to-source diversity from epoch-to-epoch variability, so we do not interpret those components individually. The simulations use the corresponding root-mean-square predictive width $\sigop=[\expt(\sigma_R^2)]^{1/2}$. Standard Markov-chain convergence diagnostics are satisfactory; for completeness, these are reported in Appendix~\ref{app:archive}. The resulting predictive width, rather than the formal uncertainty on the fitted slopes, sets the practical sensitivity to a high-mass radio excess.}

\subsection{Proxy-Mass Archive Analysis}\label{sec:hierarchy}

\rev{The 72-BCG archive is fundamentally different from the calibration sample: its black hole masses were inferred from $K$-band host-galaxy luminosity rather than measured dynamically, and 30 of the 72 nuclei have only upper limits in hard X-rays. We therefore ask a narrower question: can the reported radio excess be distinguished from an error in the proxy mass scale? The model allows the true black hole masses to differ from their $K$-band estimates, including a possible zero-point shift $\delta_K$ and a mass-dependent stretch, while simultaneously fitting the mass--X-ray relation and a mean BCG radio displacement $\Delta_R$. We also allow this heterogeneous archive to have additional radio dispersion beyond that measured in the direct-mass calibration sample. X-ray upper limits are retained as limits rather than discarded or converted into artificial detections.}

\rev{A second concern is that the archive was assembled from a mixture of targeted observations and archival availability, so its probability of including a radio-faint BCG is unknown. We cannot reconstruct that selection function from the published sample alone. Instead, we repeat the inference under a family of plausible radio-flux selection curves. Each curve has a characteristic flux $f_0$ and a width $w$ that controls how gradually the inclusion probability changes from low to high values; the exact form is given in Appendix~\ref{app:archive}. We then examine whether the inferred radio excess survives these plausible changes in sample selection. This deliberately avoids presenting a single ``selection-corrected'' offset that the available data cannot justify.}

\subsection{High-Mass Break Test}\label{sec:fisher}

\rev{The observational question is not simply whether the most massive objects have positive radio residuals. A genuine high-mass transition must produce a pattern that cannot be reproduced by an ordinary change in the FP zero point, a modest change in the mass slope, or another measured astrophysical property. Finite mass uncertainties also blur an intrinsically sharp transition: for object $i$, a step at $\mu_{\rm brk}$ is represented by $g_i=P(\mu_i>\mu_{\rm brk}\mid\widehat\mu_i)$. Thus $g_i$ approaches zero for an object securely below the trial break, approaches unity for one securely above it, and takes an intermediate value when the dynamical-mass uncertainty straddles the transition. Over the approximately $1\dex$ mass span of the direct-mass BCGs, much of this smooth pattern can be reproduced by shifting the FP normalization and slightly changing its mass dependence. That part cannot be claimed as evidence for a new high-mass regime.}

\rev{We therefore isolate only the part of the proposed break signature that remains after the ordinary FP trends and other measured source properties have been allowed to adjust. Let the columns of $A$ describe these alternative trends. The matrix $C$ summarizes the expected uncertainties in the radio residuals and the correlations among them, including the measured FP width, propagated mass uncertainty, measurement noise, and uncertainty in the common calibration coefficients. The derivation can be written compactly as $y=Aq+\Delta g+\epsilon$, with $\epsilon\sim\mathcal N(0,C)$, where $q$ contains the ordinary FP adjustments and $\Delta$ is the amplitude of the trial high-mass change. For a fixed $\Delta$, fitting $q$ removes the part of $g$ that can be reproduced by the columns of $A$, leaving $g_\perp$. After these ordinary trends have been fitted, the likelihood for $\Delta$ depends on the remaining pattern $y_\perp-\Delta g_\perp$. Its curvature with respect to $\Delta$, which sets the expected precision of the fitted break amplitude, is $g_\perp^{\mathsf T}C^{-1}g_\perp$. This gives}
\begin{equation}
\Istep=g_\perp^\trans C^{-1}g_\perp,\qquad g_\perp=g-A(A^\trans C^{-1}A)^{+}A^\trans C^{-1}g,
\label{eq:istep}
\end{equation}
\rev{where the superscript $+$ denotes the Moore--Penrose pseudoinverse, a generalized matrix inverse that remains well defined when two fitted trends contain the same information. This situation occurs, for example, if an environmental indicator is constant across an all-BCG sample and is therefore indistinguishable from the overall zero point. In practice, a singular-value decomposition is used simply to identify and remove such duplicated directions before the break test is evaluated. The expected $1\sigma$ uncertainty on the recovered break amplitude is then $\sigma(\widehat\Delta)\simeq\Istep^{-1/2}$. To put this sensitivity on an intuitive scale, we compare it with the information the same sample would contain for a uniform radio shift affecting every object equally, represented by the $N$-component vector $\mathbf{1}$,}
\begin{equation}
\Rinfo(\mu_{\rm brk})=\frac{\Istep}{\mathbf 1^\trans C^{-1}\mathbf 1}.
\label{eq:rinfo}
\end{equation}

\rev{The denominator in Equation~(\ref{eq:rinfo}) is an idealized reference: it is the information that the same sample would provide for a coherent radio shift affecting every object equally, not a separate physical model of the FP. Here $R_{\rm info}=1$ would mean that a localized high-mass transition is as easy to measure as this uniform shift, whereas $R_{\rm info}\ll1$ means that most of the apparent transition can be absorbed by ordinary FP trends or other measured effects. The sharp additive step is used as a deliberately simple benchmark because both its amplitude and mass location can be tested. A non-detection of this benchmark does not exclude a gradual transition, a change in FP slope, a change in the radio--X-ray coupling, or a change in the intrinsic dispersion. Those physically distinct possibilities are considered explicitly below.}

\subsection{Mock-Catalog Validation}\label{sec:sbc}
\rev{Because the existing BCG archive has an unknown selection history and because the high-mass experiment must search over more than one possible break mass, we test the full analysis on simulated catalogs before interpreting its uncertainties or detection rates. The aim is straightforward: when the input astrophysical signal is known, does the analysis recover it with the accuracy that its quoted intervals imply?}

\rev{For the proxy-mass archive, we generate mock BCG samples with a fixed known radio offset, refit each sample in exactly the same way as the real archive, and compare the recovered offset with the value inserted into the simulation. This directly tests the archive measurement procedure in the range relevant to previous reports of radio-bright BCGs. It shows both the mean bias and how often the quoted $68\%$ and $95\%$ intervals actually contain the true injected offset. For the future direct-mass surveys, we search over $9.65\le\mu_{\rm brk}\le10.35$ rather than assuming that the break mass is known in advance. The detection threshold is calibrated using independent catalogs containing no break, so that the probability of obtaining a false detection anywhere in this mass scan is limited to $1\%$. We count a mock survey as successful only when it both detects the inserted transition and recovers its location to within $0.10\dex$. This is intentionally stricter than asking only whether some departure from the FP is present.}

\section{Results}\label{sec:results}

\begin{figure*}[t]
\centering
\includegraphics[width=\textwidth]{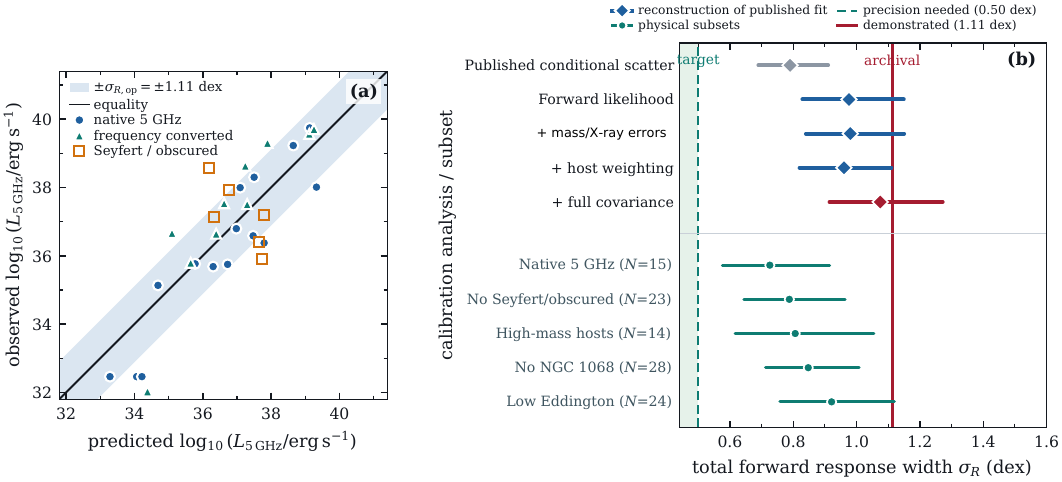}
\caption{\rev{\textit{Left:} Observed 5-GHz luminosities compared with predictions obtained by refitting the FP after withholding each host in turn. The shaded band marks $\pm\sigop$, and the symbols distinguish native 5-GHz measurements, frequency-converted data, and Seyfert or obscured systems. \textit{Right:} Forward predictive widths for successive calibration choices and several physically motivated subsets. The published $0.79\dex$ conditional scatter is shown separately because it represents a different quantity. The dashed and solid vertical lines mark the $0.50\dex$ design target and the $1.11\dex$ predictive width adopted for the mock surveys.}}
\label{fig:width}
\end{figure*}

\subsection{FP Predictive Width}\label{sec:width}

We first ask how precisely the forward relation predicts radio luminosity when fit to the only sample with direct dynamical masses. Fitting Equation~\ref{eq:fp} to the 29 direct-mass hosts gives slopes of  $b_X=\CalBXMed^{+\CalBXPlus}_{-\CalBXMinus}$ and $b_M=\CalBMMed^{+\CalBMPlus}_{-\CalBMMinus}$, consistent with previous radio-direction calibrations \citep{Gultekin2019,Plotkin2012}, but a dispersion
\begin{equation}
\sigma_R=\CalWidthMed^{+\CalWidthPlus}_{-\CalWidthMinus}\dex,
\end{equation}
with $\sigop=\OperatingWidth\dex$. The predicted and observed 5\,GHz luminosities agree only to within about an order of magnitude, across five decades in luminosity; \rev{the predictions obtained by withholding each host in turn track the equality line only loosely}, and the $\pm\sigop$ band spans more than two orders of magnitude in observed luminosity at fixed predicted luminosity (shown in Figure~\ref{fig:width}(a)). Native 5-GHz measurements and frequency-converted data scatter about the relation by comparable amounts, so the width cannot be attributed primarily to frequency conversion. NGC~1068 and NGC~0221 lie furthest from the relation, but as the subset tests below confirm, neither single object drives the result.

It is worth distinguishing this total predictive width from a smaller number already in the literature. The published $0.79\dex$ conditional scatter and the $1.07\dex$ total predictive width answer different questions and should not be interpreted as successive corrections to one common estimated quantity. \rev{The forward predictive width changes only modestly as measurement uncertainties in mass and X-ray luminosity, repeated observations of the same host, and the common distance dependence of the observables are successively included, and it remains close to the measured value in every case (Figure~\ref{fig:width}b).} What we report throughout is the total predictive width of the dataset as practised, the quantity relevant to predicting a new object's radio luminosity, rather than the conditional scatter of a fitted population model.

Removing candidates one at a time narrows the width but never enough: excluding NGC~1068 gives $0.85\dex$, excluding all flagged Seyfert or obscured systems $0.79\dex$, restricting to native 5-GHz data $0.73\dex$, and restricting to the high-mass half $0.81\dex$. These subsets are small and their intervals overlap the full-sample result. Simultaneity cannot be tested at all: only one host has radio and X-ray data within 30 days, and only four within a year. The inference is not that a single contaminant explains the width, but that \rev{the existing observational datasets superpose several known sources of variance that no archival cut can separate}. This $\sim\!1\dex$ width, not the narrower conditional scatter, is the precision against which any candidate high-mass offset or break must be judged.

\begin{figure*}[t]
\centering
\includegraphics[width=\textwidth]{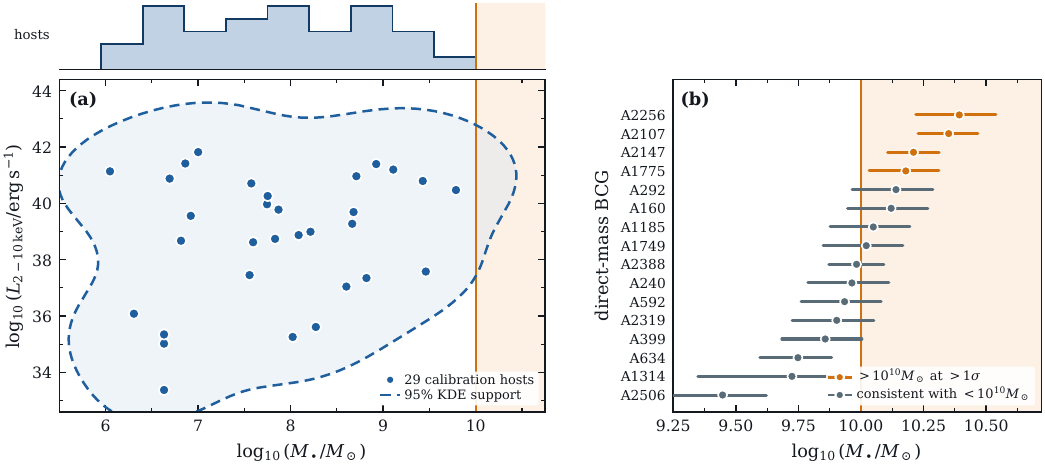}
\caption{\rev{\textit{Left:} The 29 calibration hosts in the black-hole-mass--X-ray-luminosity plane. The dashed curve is not an ellipse: it is the $95\%$ contour of a two-dimensional smoothed density estimate of the calibration sample. It is shown only to visualize the region of $(M_\bullet,L_X)$ space represented by those 29 hosts. The vertical line and shaded band mark $M_\bullet=10^{10}\,\Msun$, and the upper histogram shows the mass coverage. \textit{Right:} The 16 direct-mass BCGs ordered by mass. The plotted two-sided Gaussian mass intervals include the adopted $0.10\dex$ systematic floor. The vertical line again marks $10^{10}\,\Msun$; objects whose working mass distribution places them above this threshold at greater than $1\sigma$ are shown in orange.}}

\label{fig:support}
\end{figure*}

\subsection{High-Mass Extrapolation}\label{sec:support}

The predictive width established in the previous section is only half the problem, as we also need to understand whether the calibration sample covers the mass range in which a break is being sought. The 29 calibration hosts are concentrated well below $10^{9}\,\Msun$, with their mass distribution falling off sharply toward higher mass and only a handful of hosts approaching $10^{10}\,\Msun$ (Figure~\ref{fig:support}a). The direct-mass BCGs, by contrast, sit at or beyond this sample's high-mass edge, and two of the sixteen fall outside the empirical $95\%$ calibration-support region entirely, meaning the joint mass--X-ray combination for these objects was never represented, even sparsely, among the calibration hosts. A relation tested by withholding each calibration host in turn within the measured domain is therefore being asked to extrapolate into a regime where it has barely been measured, rather than interpolate within a domain it has already tested.

The direct-mass BCGs themselves span a narrow range close to the proposed break, which sharpens this problem rather than resolving it. Ranking all sixteen by mass, four objects, A2256, A2107, A2147, and A1775, exceed $10^{10}\,\Msun$ at greater than $1\sigma$, while the remaining twelve are consistent with lying below this threshold; several of the twelve, however, sit close enough to the boundary that their classification depends on the adopted mass-uncertainty distribution rather than being secure at high significance (Figure~\ref{fig:support}b). The sample therefore straddles the proposed break in the sense required for a transition test, but with the calibration barely extending into this mass range at all, that straddling reflects a favourable target geometry rather than a favourable measurement precision. This is a strong argument for measuring the response in the target regime rather than importing a relation calibrated almost entirely at lower mass.

\begin{figure*}[t]
\centering
\includegraphics[width=\textwidth]{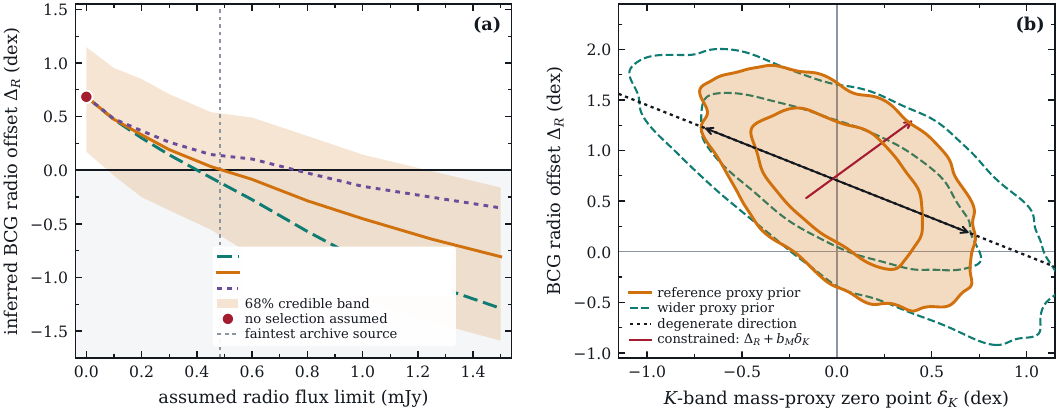}
\caption{\rev{\textit{Left:} The inferred BCG radio offset $\Delta_R$ as the assumed faint-end radio selection is varied. Dashed, solid, and dotted curves correspond to three values of the width $w$ of the selection transition: small $w$ represents a relatively abrupt change from low to high inclusion probability with radio flux, while larger $w$ represents a more gradual change. The shaded band gives the $68\%$ credible interval for the reference curve, the point at zero flux limit gives the result when no radio selection is imposed, and the vertical dotted line marks the faintest source present in the archive. \textit{Right:} Joint posterior for the radio offset $\Delta_R$ and the $K$-band mass-proxy zero point $\delta_K$, under the reference proxy prior (solid orange contours) and a wider proxy prior (dashed teal contours). Both parameters shift the predicted radio residual in nearly the same way, producing the elongated direction $\Delta_R=-b_M\delta_K$ (dotted line). The red arrow marks the combination $\Delta_R+b_M\delta_K$ that is constrained substantially better than either term separately.}}
\label{fig:archive}
\end{figure*}

\subsection{Proxy-Mass Archive Constraints}\label{sec:archive}

Having established that the calibration barely reaches the target mass range, we now ask what the proxy-based mass catalogue itself can say about a BCG radio excess. Taken at face value, assuming the 72 BCGs were included without regard to radio flux, the archive does favour a positive offset, $\Delta_R=\HistRawMed$ with a $68\%$ interval $[\HistRawLo,\HistRawHi]\dex$ (Table~\ref{tab:archive}). Three effects prevent it from being read as evidence for ultramassive black holes.

\begin{deluxetable}{lccl}[!htb]
\tablecaption{What the 72-BCG proxy-mass archive constrains.\label{tab:archive}}
\tablehead{\colhead{Quantity} & \colhead{Median} & \colhead{$68\%$ interval} & \colhead{Status}}
\startdata
Radio offset & 0.68 & $0.17$--$1.15$ & depends on archive model \\
$\Delta_R+b_M\delta_K$ & 0.71 & $0.28$--$1.10$ & well-constrained combination \\
\enddata
\tablecomments{\rev{In the mock-catalog validation, nominal $68\%$ and $95\%$ intervals contain the known injected offset in only about $50\%$ and $85\%$ of realizations. Both rows assume no radio-flux selection.}}
\end{deluxetable}

The mass proxy and the radio offset are nearly degenerate: if the $K$-band masses are systematically low by $\delta_K$, the predicted radio luminosity falls by $b_M\delta_K$ and the sample appears radio-bright by exactly that amount. The data therefore constrain
\begin{equation}
\Delta_R+b_M\,\delta_K = \HistInvariantMed^{\,+0.39}_{\,-0.43}\dex
\label{eq:invariant}
\end{equation}
far better than either term individually. The joint posterior for $\Delta_R$ and $\delta_K$, shown under both a reference proxy prior and a substantially wider alternative, is elongated almost exactly along the direction $\Delta_R=-b_M\delta_K$ in both cases; the two contour sets differ considerably in overall extent but share essentially the same orientation, so widening the proxy prior enlarges the allowed region without breaking the degeneracy (Figure~\ref{fig:archive}b). The red arrow marking the identified combination of Equation~(\ref{eq:invariant}) cuts across both contour sets at a shallow angle, confirming that this sum, and not either parameter separately, is what the archive actually pins down. \rev{This is an observational degeneracy:} a positive value of the combination is consistent either with a genuine radio offset or with systematically underestimated proxy masses. Only independently measured masses and matched proxy photometry can separate the two terms. 


The sign of the offset is also sensitive to unknown selection. \rev{As the assumed faint-end radio flux limit is raised from zero, the inferred offset $\Delta_R$ declines steadily for all three adopted widths of the selection transition, tracing out nearly parallel curves that diverge only at the highest flux limits;} the point at zero flux limit, corresponding to no selection at all, sits well above the curves, illustrating how much of the raw offset is attributable to the assumption of complete inclusion (Figure~\ref{fig:archive}a). The offset crosses zero at a flux limit of only a few tenths of a mJy, below the flux of the faintest source actually present in the archive, so the archive's own faintest detection cannot rule out a selection function strong enough to erase the offset entirely. Because the parent sample and \rev{the unobserved faint-end radio-flux distribution} are unavailable, this demonstrates sensitivity to the assumed inclusion model; it does not determine the true population offset or its sign.

\rev{Finally, the quoted interval widths are too optimistic. In the mock-catalog tests, nominal $68\%$ and $95\%$ intervals contain the known injected offset only about $50\%$ and $85\%$ of the time, respectively, and the recovered offset has an approximately constant $+0.10\dex$ bias (Figure~\ref{fig:calibration}). Because neither the measured bias nor these overly narrow intervals have a unique correction without knowing the true archive selection function, we do not apply an ad hoc recalibration. Taken together, the proxy-mass degeneracy, selection sensitivity, and validation failure mean that the raw archive offset cannot be promoted to a population-level measurement of a high-mass FP departure.}

\begin{figure*}[t]
\centering
\includegraphics[width=\textwidth]{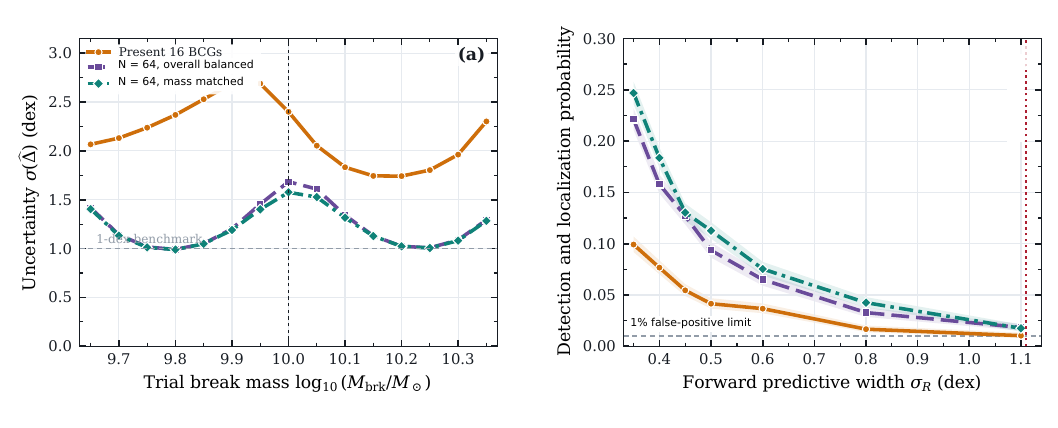}
\caption{\rev{\textit{Left:} Expected $1\sigma$ uncertainty on the amplitude of a high-mass transition as the assumed break mass is moved across the direct-mass BCG sample, for the present 16 BCGs (dash-dot) and two representative $N=64$ surveys (overall balanced, dashed; mass-matched, dotted). The horizontal dashed line marks the 1-dex transition used as the benchmark experiment. \textit{Right:} Fraction of mock surveys that both detect this 1-dex transition and recover its mass location to within $0.10\dex$, as a function of the achieved FP predictive width $\sigma_R$. The vertical line marks the width measured from the present calibration. The false-positive threshold is calibrated so that only $1\%$ of no-break mock catalogs produce a spurious detection anywhere in the full scanned mass interval. Shaded bands are $95\%$ Monte Carlo intervals.}}
\label{fig:detect}
\end{figure*}

\subsection{High-Mass Break Detectability}\label{sec:info}

\rev{The previous sections establish that the existing FP prediction is broad and that the proxy-based BCG radio offset is not uniquely interpretable. We now ask directly how large the uncertainty would be on a genuine high-mass transition. Before quoting the detection probabilities, we define the two $N=64$ survey layouts used in the simulations. Both contain the same black-hole masses, mass uncertainties, and X-ray luminosities. In the ``overall-balanced'' survey, the total numbers of objects in each environmental or accretion category are equal, but a category may still occur preferentially on one side of the proposed mass break. In the ``mass-matched'' survey, both environmental classes and both accretion-state classes are represented on each side of the trial break at comparable black-hole masses. The second layout is therefore designed to distinguish a mass dependence from changes associated with cluster environment or accretion state.}

\rev{For the present 16 BCGs, the expected $1\sigma$ uncertainty on a transition at $\mu_{\rm brk}=10$ is $\sigma(\widehat\Delta)=2.40\dex$. The corresponding uncertainties are $1.68\dex$ for the overall-balanced $N=64$ survey and $1.58\dex$ for the mass-matched $N=64$ survey. Matching environment and accretion state across mass therefore helps separate competing physical effects, but by itself it does not make a 1-dex mass break readily measurable. This uncertainty is not flat across trial break masses: as Figure~\ref{fig:detect} shows, $\sigma(\widehat\Delta)$ peaks for all three samples near $\mu_{\rm brk}=10$, precisely the mass at which a break is being sought. Near this mass, neither $N=64$ design reaches the $1\dex$ benchmark uncertainty; values near $1\dex$ occur only farther from the trial mass of primary interest. The same limitation is reflected by $R_{\rm info}$: for the present sample it remains only about $1$--$19\%$ of the sensitivity to a simple uniform radio shift across the range of mass-uncertainty assumptions tested in Appendix~\ref{app:calibration}.}

At the demonstrated width $\sigma_R=\OperatingWidth\dex$, joint detection and $0.10\dex$ localization of a 1-dex break occur in $1.0\%$ of present-sample catalogs, $1.8\%$ of overall-balanced $N=64$ catalogs, and $1.8\%$ of mass-matched $N=64$ catalogs. At $\sigma_R=0.40\dex$, the corresponding probabilities are $7.7\%$, $15.8\%$, and $18.4\%$. As Figure~\ref{fig:detect}(b) shows, this joint success probability rises steeply as $\sigma_R$ narrows below the measured value, but even at the narrowest widths considered the present 16-BCG sample remains well short of a reliable detection, underscoring that sample size, not just precision, limits this design. \rev{For the mass-matched configuration at $\sigma_R=0.40\dex$, the mock-survey sample-size calculation first exceeds a $70\%$ joint-success target at approximately $N=320$, where the success probability is $71.6\%$ with a $95\%$ Monte Carlo interval of $70.3$--$72.8\%$.} \rev{These numbers are conditional on a sharp 1-dex transition, the adopted mass-error distributions, and the specified set of alternative trends. The quoted $1\%$ false-positive level applies to the entire search over trial break masses, not to each mass separately: among mock catalogs containing no break, only about $1\%$ are allowed to produce a spurious detection somewhere in the scanned mass interval.}

\begin{deluxetable}{lccc}[!htb]
\tablecaption{Detection performance of the mock survey designs.\label{tab:designs}}
\tablehead{\colhead{Design} & \colhead{$\sigma(\widehat\Delta)$} & \multicolumn{2}{c}{Joint success} \\
\colhead{} & \colhead{at $10^{10}\Msun$} & \colhead{$\sigma_R=1.11$} & \colhead{$\sigma_R=0.40$}}
\startdata
Present 16 BCGs & 2.40 dex & 1.0\% & 7.7\% \\
$N=64$, overall balanced & 1.68 dex & 1.8\% & 15.8\% \\
$N=64$, mass-matched & 1.58 dex & 1.8\% & 18.4\% \\
\enddata
\tablecomments{\rev{Joint success requires both detection and localization of a sharp 1-dex transition to within $0.10\dex$. The break location is searched over $9.65\le\mu_{\rm brk}\le10.35$, and the detection threshold is calibrated so that a no-break catalog has only a $1\%$ probability of producing a false detection anywhere in this full mass scan.}}
\end{deluxetable}

\subsection{Physical Origin of Radio Residuals}\label{sec:attribution}

\begin{figure*}[t]
\centering
\includegraphics[width=\textwidth]{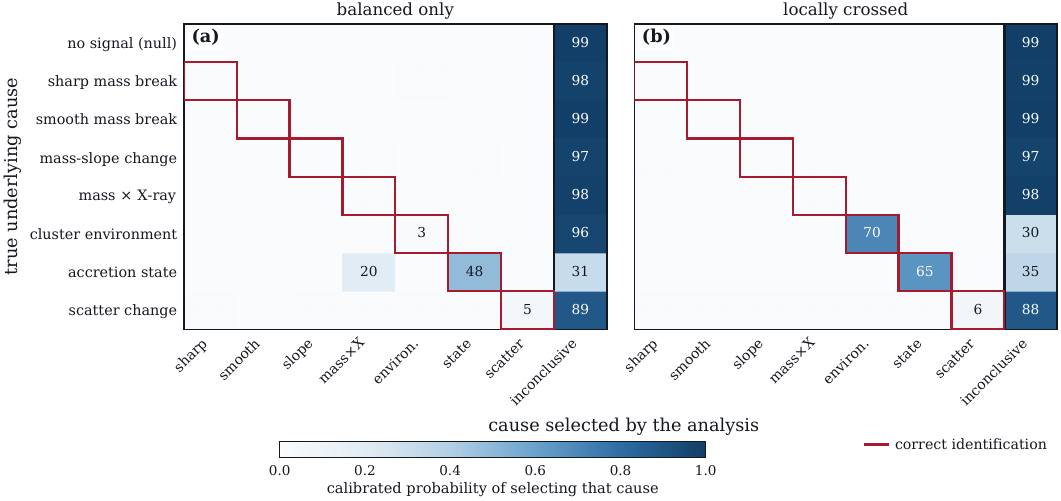}
\caption{\rev{Physical-origin test for two $N=64$ mock surveys at the demonstrated width $\sigma_R=\OperatingWidth\dex$. Rows give the physical effect inserted into the mock catalog; columns give the effect selected by the analysis, with the final column denoting an inconclusive result. Red boxes mark correct identifications. Both survey designs contain the same masses, mass uncertainties, X-ray luminosities, and the same total numbers of objects in each environmental and accretion-state category. They differ only in where those categories fall in mass. \textit{Left:} in the overall-balanced design, equal overall category counts do not prevent environment or accretion state from remaining correlated with mass near the proposed break. \textit{Right:} in the mass-matched design, each environmental and accretion-state category is represented on both sides of the trial break within comparable mass intervals. The detection threshold is calibrated so that the probability of selecting any spurious physical cause when none is present is $1\%$.}}
\label{fig:attribution}
\end{figure*}

\rev{Even if future observations reveal a systematic radio residual, its physical origin would remain ambiguous unless mass is separated from the other properties that change across the BCG population. Figure~\ref{fig:attribution} therefore tests a set of deliberately distinct astrophysical possibilities. The ``sharp mass break'' represents an abrupt change in radio normalization above a characteristic black hole mass, while the ``smooth mass break'' represents a gradual transition. The ``mass-slope change'' and ``mass $\times$ X-ray'' cases represent changes in the mass dependence of the FP and in the coupling between the radio-emitting jet and X-ray-emitting accretion flow. Such changes could arise if jet efficiency or magnetic-flux saturation changes systematically toward the highest masses \citep{Tchekhovskoy2011}. The ``cluster environment'' case represents a radio displacement associated with the thermodynamic state of the surrounding intracluster medium, motivated by the known connection between BCG radio activity, cooling atmospheres, and feedback \citep{McNamaraNulsen2007,Cavagnolo2010,Hogan2015b}. The ``accretion state'' case represents a change associated with Eddington ratio or nuclear accretion regime \citep{YuanCui2005,Kording2006,Abbassi2026}. Finally, the ``scatter change'' represents a population in which the mean FP remains similar but the object-to-object dispersion changes, as could occur if more than one accretion or jet mode contributes at the highest masses.}

\rev{Each mock catalog is generated with one of these effects, or with no additional effect, and is then analysed with each of the same candidate residual patterns while carrying the dynamical-mass uncertainties through the fit (Appendix~\ref{app:methods}). A physical cause is reported only when its fitted residual pattern exceeds a detection threshold established from independent mock catalogs containing no additional effect; otherwise the result is labelled inconclusive. The threshold is set so that only $1\%$ of no-effect catalogs falsely select any physical cause across the complete set of tests. This mock-catalog comparison asks a more demanding question than whether the FP residuals are non-zero: it asks whether their pattern contains enough information to distinguish the physical mechanism that produced them.}

\rev{At the demonstrated width, signatures tied directly to black hole mass cannot be reliably distinguished. Sharp and smooth transitions, changes in the mass slope, and changes in the mass--X-ray coupling are labelled inconclusive in at least $97\%$ of the simulations. A change in the intrinsic scatter is correctly recognized in only about $5$--$6\%$. The reason is astrophysical as well as statistical: over the narrow high-mass range, these different forms of mass dependence produce radio-residual patterns that are small compared with the present object-to-object dispersion.}

\rev{The situation improves for effects that can be deliberately crossed with mass. In the mass-matched survey, a cluster-environment effect is correctly identified in $70\%$ of realizations, compared with only $3\%$ in the overall-balanced survey, while an accretion-state effect is recovered in $65\%$ compared with $48\%$. The gain comes from placing objects with different environments and accretion states at comparable black hole masses on both sides of the trial transition. In the overall-balanced survey, $20\%$ of simulated accretion-state signals are instead assigned to the mass--X-ray alternative. This confusion is physically understandable because the Eddington ratio satisfies $\lambda_{\rm Edd}\propto L_{\rm bol}/M_\bullet$ and therefore couples luminosity to black-hole mass; if accretion state is not matched across mass, a state-dependent radio residual can resemble an explicit mass--X-ray dependence. The experiment can then ask whether the residual follows mass itself, the cluster atmosphere, or the accretion regime. When the simulated population contains no additional physical effect, no cause is selected in $99\%$ of realizations, as required by the adopted false-positive calibration.}

\section{Discussion}\label{sec:discussion}

\subsection{Origin of the FP Scatter}\label{sec:physical}

The width found in Section~\ref{sec:width} is physically expected for a heterogeneous centimetre compilation, and it is not a single effect. Centimetre beams mix an unresolved core with jets and relic plasma whose luminosity depends on source age, confinement and the cluster atmosphere \citep{McNamaraNulsen2007,Hogan2015b}. Radio and X-ray epochs are typically separated by years, so variable nuclei are sampled at unrelated phases. Hard X-ray luminosities can be suppressed by absorption or inflated by thermal intracluster emission and unresolved X-ray binary populations \citep{Gilfanov2004}. \rev{At low Eddington ratio, the coupling among accretion rate, coronal X-ray emission, and compact-jet radio emission is itself state dependent \citep{YuanCui2005,Kording2006,Abbassi2026}.} And host-to-host differences in magnetic flux, spin and jet efficiency contribute genuine physical diversity \citep{Tchekhovskoy2011}.

The fitted slopes remain compatible with previous calibrations, but that compatibility should not be overinterpreted: a common slope can coexist with a state-dependent intercept or variance, and a high-mass sample is in practice also a low-Eddington sample. Any future census must therefore cross mass with an independent accretion-state diagnostic rather than treating high mass as a proxy for accretion regime.

\subsection{Offset versus Break}\label{sec:offset}

\rev{The earlier proxy-mass BCG result and the direct-mass break experiment address different astrophysical questions.} A coherent BCG offset says that these galaxies as a class are radio-loud relative to the FP; it can be supported by every object in a sample, and it is sensitive to radio selection and to the mass-proxy zero point. A high-mass break says the relation itself changes character above a particular mass; it must be localised in mass and survive removal of ordinary trends and other measured source properties. Equation~(\ref{eq:invariant}) shows why the first cannot be converted into the second: even a securely measured positive value of $\Delta_R+b_M\delta_K$ would not distinguish an ultramassive mass correction from resolution effects, environment, accretion state, or selection. Conversely, the direct-mass test can look for a localized break regardless of the sign of the earlier proxy-mass offset. Combining the two as one measurement would reintroduce precisely the circularity that direct masses were obtained to remove.

\subsection{Observing Strategy}\label{sec:program}

The results above point to a clear operational conclusion: precision must be established before any population study is attempted. We therefore propose a two-stage program, beginning with a dedicated precision experiment rather than a search for a break. The first stage targets 32--48 galaxies with secure direct dynamical masses spanning the full mass range, and measures their nuclei under one common protocol. \rev{The currently assembled non-overlapping pool contains 45 unique systems, so the upper end of this range is a sample-size requirement rather than a ready-made target list; target selection should be fixed from mass provenance and observability before the new radio measurements are examined.} The radio observable is a model-fitted unresolved core flux at a common rest-frame 5 GHz, recovered over a matched baseline range and reported at a common physical scale. We adopt 100\,pc as the reference: at the BCG distances this is $0.07$--$0.19''$ (Table~\ref{tab:bcgs}), within reach of existing centimetre interferometry, with 50- and 200-pc reductions as predeclared sensitivities. The X-ray observable is the intrinsic nuclear $2$--$10$\,keV luminosity from a predeclared aperture and local background, fitted with a Poisson count model for source and local-background regions, including a thermal cluster component, absorption, and explicit Compton-thick treatment \citep{ChandraPOG}, with low-count sources retained as upper limits. Radio and X-ray epochs should be matched within 30 days. A compact millimetre measurement is a valuable complement, since it suppresses steep-spectrum extended emission and has already yielded a millimetre FP with low contamination \citep{Ruffa2024}.

\rev{This matched-scale requirement is physical, not merely procedural. A radio residual can be interpreted in terms of black hole mass only when the radio measurement isolates the compact core and the X-ray luminosity traces the corresponding accretion flow. Recent tests of inverse-FP applications likewise show that low Eddington ratio alone is not sufficient to establish these conditions \citep{Abbassi2026}. The common spatial scale and near-simultaneous radio/X-ray measurements proposed here are intended to reduce precisely these ambiguities.}

\begin{figure*}[t]
\centering
\includegraphics[width=\textwidth]{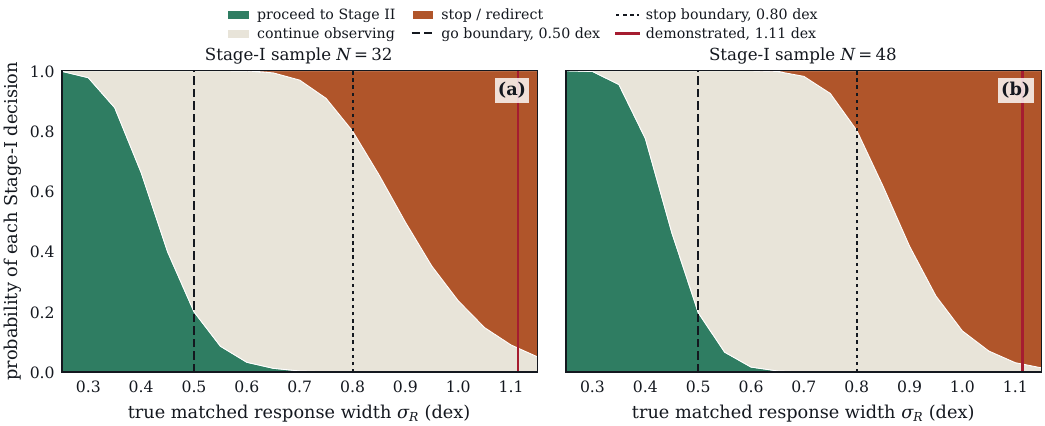}
\caption{Expected Stage-I decision probabilities, Equations~(\ref{eq:go})--(\ref{eq:stop}), assuming a $0.25\dex$ contribution from measurement uncertainty. Shaded regions give the probability of each decision as a function of the true matched response width, for Stage-I samples of $N=32$ and $N=48$. The rule deliberately leaves a broad continuous region rather than committing to an expensive census before the required precision is demonstrated.}
\label{fig:stage1}
\end{figure*}

\rev{This stage concludes with an explicit decision rule specified in advance, with thresholds set by the detection-probability calculation of Section~\ref{sec:info}. Let $D_1$ denote the Stage-I measurements:}
\begin{align}
\text{proceed:}\quad & P(\sigma_R<0.50\dex\mid D_1)\ge0.80,\label{eq:go}\\
\text{stop or redirect:}\quad & P(\sigma_R>0.80\dex\mid D_1)\ge0.80,\label{eq:stop}
\end{align}
Posterior probabilities falling between these two thresholds are classified as continue, indicating that Stage I should proceed with additional targets before a decision is made. If the matched catalogue truly achieves $0.40\dex$, this rule allows Stage I to proceed with probability $\SeqReachThirtyTwo\%$ for $N=32$ and $\SeqReachFortyEight\%$ for $N=48$; if the width is really $1.0\dex$, the program correctly stops in $76\%$ and $86\%$ of cases; and at today's archival width the probability of proceeding is effectively zero (Figure~\ref{fig:stage1}). Folding both stages together, the overall probability of reaching the next stage and obtaining a decisive result is $\SeqOverallThirtyTwo\%$ ($N=32$) and $\SeqOverallFortyEight\%$ ($N=48$) for a true $0.40\dex$ response. \rev{These probabilities assume that the pilot dispersion is not itself broadened by a real but as-yet undetected change in the FP. If the mean radio relation changes across the sampled mass range, fitting all objects with a single relation would incorrectly attribute part of that change to random scatter. Equations~(\ref{eq:go})--(\ref{eq:stop}) should therefore be regarded as planning thresholds rather than a final hard gate; the pilot analysis should fit any mass-dependent change in the mean relation together with the dispersion within each regime.}

\rev{Stage II is justified only if the pilot demonstrates that substantially narrower nuclear residuals are achievable. Its sample must cover both sides of $10^{10}\,\Msun$ with BCG and non-BCG hosts and, within each local mass range, must include nuclei spanning comparable accretion regimes. In practical terms, black hole mass should not be allowed to stand in for either cluster environment or Eddington ratio. We propose the central intracluster entropy $K_0$ as the primary environmental diagnostic, used continuously with the conventional cool-core division as a robustness check \citep{Cavagnolo2010}. The primary accretion diagnostic is the nuclear Eddington ratio inferred from the hard X-ray luminosity and the dynamical mass. Uncertainty in these quantities should be carried into the analysis rather than assigning every galaxy to an exact environmental or accretion-state category. This matched design is what gives any high-mass radio residual a realistic chance of being associated with black hole mass itself rather than with the surrounding atmosphere or accretion regime.}

\subsection{Future Facilities}\label{sec:future}
The first stage of the measurement is within reach now, but forthcoming facilities would transform it. The sensitivity and resolution of the ngVLA would allow the unresolved core to be isolated at tens of parsecs rather than $\sim\!100$\,pc in these systems, directly attacking the dominant contributor to the width \citep{ngVLA2025}, while SKA-era surveys will supply the homogeneous low-frequency context needed to separate core from relic emission \citep{SKAO2025}. On the X-ray side, the confusion between a faint nucleus and the surrounding intracluster medium is the binding limitation, and high-resolution spectroscopic imaging of the kind proposed for next-generation X-ray observatories would substantially reduce it \citep{Lynx2019,Cruise2025}. Our forecasts are deliberately expressed in terms of the achieved response width rather than a particular instrument, so they can be read directly against whatever combination of facilities becomes available, using the width-to-detection-probability mapping already established in Figure~\ref{fig:detect}.

\subsection{Limitations}\label{sec:limits}

Our forecasts apply to the declared experiment and inherit its assumptions. The calibration is conditioned on nuclei with usable radio and X-ray detections, and is heterogeneous by construction. The BCG mass-error distributions approximate the published intervals rather than reconstructing orbit-model likelihoods, and should be replaced when those become available; the systematic range they induce is quantified in Appendix~\ref{app:calibration} rather than propagated into the headline results above. \rev{The proposed $N=64$ catalogue is an illustrative feasibility design, not a demonstrated target list.} \rev{Finally, the staged program assumes that the measurement control achieved in Stage I carries over to Stage II.} The future-survey simulation can be rerun end to end, but the original samplers for the empirical calibration and proxy-mass archive model are not included in the companion archive. The sample-size calculation also does not model selection into the direct-mass census; $N\simeq320$ is therefore a conditional feasibility boundary rather than a demonstrated target list. None of these assumptions weakens the central point: the existing archive cannot resolve a mass break, and any future attempt to do so must first demonstrate the required precision before committing to a large population study.

\section{Conclusions}\label{sec:conclusions}

\rev{We have used direct dynamical black hole masses, the observed forward scatter of the radio--X-ray Fundamental Plane, and validated mock-survey experiments to ask whether a change in the FP above $10^{10}\,\Msun$ can be recognized with present data and what observations would be required to establish one.} Our main results are summarised as follows:

\begin{enumerate}

\item Rebuilt from 29 hosts with direct dynamical masses, the centimetre/X-ray Fundamental Plane predicts radio luminosity with a total forward width of $\CalWidthMed^{+\CalWidthPlus}_{-\CalWidthMinus}\dex$. This is the predictive dispersion of the heterogeneous protocol and is not directly comparable to a conditional intrinsic-scatter estimate.

\item The proxy-mass 72-BCG archive constrains the combination $\Delta_R+b_M\delta_K$ far better than either a genuine radio offset or the $K$-band mass-proxy zero point separately. Its sign is sensitive to assumed selection functions, and \rev{mock-catalog validation shows both a measurable bias and uncertainty intervals that contain the true input less often than their nominal confidence levels imply}; no corrected population interval is therefore claimed.

\item At the demonstrated width, joint detection and localisation of a sharp 1-dex transition remain at the $1$--$2\%$ level for both the present sample and the $N=64$ designs. Even at $\sigma_R=0.40\dex$, the mass-matched $N=64$ design reaches only $18.4\%$; \rev{for the mass-matched configuration at this same predictive width, approximately $N=320$ systems are required to exceed a $70\%$ joint detection-and-localization probability under the adopted model.}

\item Local crossing of environment and accretion state against mass remains essential for interpretation, but it improves sharp-break power only modestly. A null result for the step benchmark would not exclude smoother transitions, slope changes, or changes in residual variance.

\item \rev{We therefore recommend a matched-scale, quasi-simultaneous pilot using 32--48 direct-mass nuclei. The pilot should measure the dispersion within the sampled mass regimes while simultaneously allowing for a possible change in the mean FP relation, rather than interpreting all excess width as random scatter. A larger census is justified only if the pilot demonstrates both the required nuclear precision and a target population in which mass, cluster environment, and accretion state can be separated observationally.}

\end{enumerate}

The direct-mass measurements that made this analysis possible are still young, and they will grow in future. As more BCGs and comparison galaxies receive dynamical mass measurements, and as the matched-scale pilot proposed here is carried out, the two limiting factors identified in this paper, calibration precision and target-sample size, can both be directly addressed rather than assumed. \rev{The companion archive accompanying this paper makes the full future-survey simulation framework available for exactly this purpose: any future direct-mass sample, or any revised measurement of the forward relation's width, can be run through the same detectability framework to update the detection forecasts presented here.} Whether the black hole Fundamental Plane truly changes character above $10^{10}\,\Msun$ therefore remains an open question; however, this analysis provides a quantified path toward answering it, using a specific precision target, a specific sample size, and a decision rule for when a larger census is justified.

\begin{acknowledgments}
R.S. and S.M. acknowledge support from the Department of Atomic Energy, Government of India, under project no. RTI4019. \rev{The authors acknowledge the use of generative AI for text editing.}
\end{acknowledgments}

\software{NumPy \citep{Harris2020}, SciPy \citep{Virtanen2020}, Matplotlib \citep{Hunter2007}, pandas \citep{McKinney2010}}

\section*{Data availability}
\rev{The companion archive contains the manuscript source, processed tables, stored posterior samples, future-survey mock catalogs, analysis and plotting code, tests, validation logs, environment files, and checksums. The future-survey simulations can be rerun end to end. The original sampling programs for the 29-host calibration and the 72-BCG proxy-mass archive model are not present, so those two empirical fits cannot be rerun from the companion archive alone.}

\appendix
\restartappendixnumbering

\section{Model Details}\label{app:methods}

\subsection{Forward Calibration and Mock Surveys}\label{app:forward}

\rev{A future survey uses one common FP calibration for all of its galaxies. Uncertainty in the fitted zero point and slopes therefore moves the predicted radio luminosities coherently rather than acting as a new independent error for every object. We represent the full covariance of the predicted radio residuals by $C$. With $\theta=(a,b_X,b_M)$ and $Z_i=(1,\,X_i-40,\,\widehat\mu_i-8)$,}
\begin{equation}
C=\operatorname{diag}\!\bigl(\sigma_R^2+b_M^2 s_{\mu,i}^2+\sigma_{{\rm meas},i}^2\bigr)+Z\,\mathrm{Cov}(\theta)\,Z^\trans ,
\label{eq:cov}
\end{equation}
\rev{Equation~(\ref{eq:cov}) is obtained by ordinary covariance propagation. Independent object-to-object contributions add on the diagonal, while uncertainty in the common FP coefficients is propagated through the linear prediction $Z\theta$, giving $\mathrm{Cov}(Z\theta)=Z\,\mathrm{Cov}(\theta)\,Z^\trans$. The diagonal term contains the object-by-object sources of dispersion: the FP predictive width, propagation of each mass uncertainty through the FP mass slope, and the remaining measurement contribution. Here $\sigma_{{\rm meas},i}$ denotes that combined object-specific observational contribution to the radio residual after the mass term has been written explicitly. The forecast does not separate this term into radio and X-ray components; any X-ray-luminosity uncertainty propagated through the FP is included in $\sigma_{{\rm meas},i}$ rather than added as a second covariance term. The second term, $Z\,\mathrm{Cov}(\theta)\,Z^\trans$, describes the uncertainty in the three calibration coefficients. Because the same three coefficients affect every galaxy, this contribution is shared across the entire catalog and has at most three independent directions. Treating it instead as independent noise for every object would incorrectly erase the fact that a change in the FP calibration moves many predicted radio luminosities together. In each no-break mock catalog we therefore draw one common calibration vector and, when the catalog is analysed, use Equation~(\ref{eq:cov}) to represent the remaining calibration uncertainty. We do not add a second independent calibration error to each galaxy; doing so would count the same uncertainty twice. This covariance $C$ is used in Equations~(\ref{eq:istep})--(\ref{eq:rinfo}).}

\rev{After subtracting the mean radio luminosity predicted by the calibration, the residual for a mock galaxy is written as}
\begin{equation}
y_i=\alpha+\beta(\widehat\mu_i-10)+(b_M+\beta)(\mu_i-\widehat\mu_i)+\Delta\,H(\mu_i-\mu_{\rm brk})+\epsilon_i ,
\label{eq:residual}
\end{equation}
\rev{Here $H(\mu_i-\mu_{\rm brk})$ is the Heaviside step function: it is zero below the trial break mass and unity above it. The coefficient $\Delta$ is therefore the added radio-luminosity displacement, in dex, for objects above the break. The terms $\alpha$ and $\beta$ allow the overall residual zero point and a smooth residual trend with mass to adjust. Because the underlying FP itself already contains the mass slope $b_M$, propagation of uncertainty in the true mass depends on $b_M+\beta$, not on $\beta$ alone. Writing $\gamma=b_M+\beta$, let $\sigma$ denote the radio-residual width, $s_\mu$ the $1\sigma$ uncertainty in log black-hole mass, $v=\sigma^2+\gamma^2s_\mu^2$, and $\tau=s_\mu\sigma/\sqrt v$. Integrating over the unmeasured true mass on each side of a sharp break then gives}
\begin{equation}
p(y\mid\widehat\mu)=\normal(y\mid\eta_0,v)\,\Phi\!\left(\frac{\mu_{\rm brk}-\widetilde\mu_0}{\tau}\right)
+\normal(y\mid\eta_1,v)\,\Phi\!\left(\frac{\widetilde\mu_1-\mu_{\rm brk}}{\tau}\right),
\end{equation}
\rev{Here $\eta_k=\alpha+\beta(\widehat\mu-10)+k\Delta$ is the radio-residual mean in branch $k=0,1$, and $\widetilde\mu_k=\widehat\mu+\gamma s_\mu^2(y-\eta_k)/v$ is the corresponding conditional mean mass entering the analytic integral. The closed form agrees with direct numerical integration to a maximum relative error below $1.1\times10^{-13}$ over 250 test cells. A Gaussian approximation is used only to estimate the fitted break location; all headline detection probabilities use the exact calculation that integrates over the mass uncertainty.}

\subsection{Proxy-Mass Archive Model}\label{app:archive}
For archive object $i$ with tabulated proxy $\mu_{K,i}$,
\begin{align}
\mu_i&\sim\normal\bigl(\mu_{K,i}+\delta_K+\eta_K(\mu_{K,i}-9.8),\,s_K^2\bigr),\\
X_i&\sim\normal\bigl(c_0+c_1(\mu_i-9.8),\,\sigma_X^2\bigr),\\
R_i&\sim\normal\bigl(38+a+b_X(X_i-40)+b_M(\mu_i-8)+\Delta_R,\ \sigma_R^2+\sigma_{\rm extra}^2\bigr).
\end{align}
Priors are weakly informative: $\Delta_R\sim\normal(0,1.5^2)$, $\delta_K\sim\normal(0,0.30^2)$, $\eta_K\sim\normal(0,0.45^2)$, $\log s_K\sim\normal(\log0.44,0.28^2)$, $c_0\sim\normal(40.6,2^2)$, $c_1\sim\normal(0,1.2^2)$, with half-normal scales of $1.2$ and $1.4\dex$ for $\sigma_{\rm extra}$ and $\sigma_X$; the calibration vector uses a multivariate Gaussian approximation to the full calibration posterior. \rev{To test sensitivity to the unknown radio selection, Figure~\ref{fig:archive}(a) assumes that the probability for a BCG to enter the archive rises smoothly with observed radio flux,}
\begin{equation}
P(S_i{=}1)=\frac{1}{1+\exp\!\left[-(\log_{10}F_{\nu,i}-\log_{10}f_0)/w\right]}.
\end{equation}
\rev{This smooth S-shaped function is written explicitly so that no additional statistical terminology is needed. The parameter $f_0$ sets the characteristic flux at which the inclusion probability is one half, while $w$ controls how gradually the probability changes with log flux. We do not claim that this is the true selection function of the proxy-mass archive; it is used only to determine whether the inferred BCG radio offset is robust to plausible preferential inclusion of radio-bright objects. The required normalization is evaluated numerically using standard Gaussian quadrature (Gauss--Hermite quadrature).}

\rev{The stored calibration fit contains four independent Markov chains and 16{,}000 retained samples, while the proxy-mass archive fit contains 12{,}000 retained samples. For completeness, convergence was checked with the rank-normalized split-$\widehat R$ diagnostic \citep{Vehtari2021}, which compares variation within and between independent chains; values close to unity indicate that the chains are sampling the same distribution. All reported parameters have $\widehat R<1.01$ and satisfy the stated effective-sample-size thresholds. Because the original sampling programs are absent from the companion archive, these diagnostics can be inspected but the two empirical fits cannot be rerun from the archive alone.}

\section{Validation and Supporting Data}\label{app:calibration}
\begin{figure*}[t]
\centering
\includegraphics[width=\textwidth]{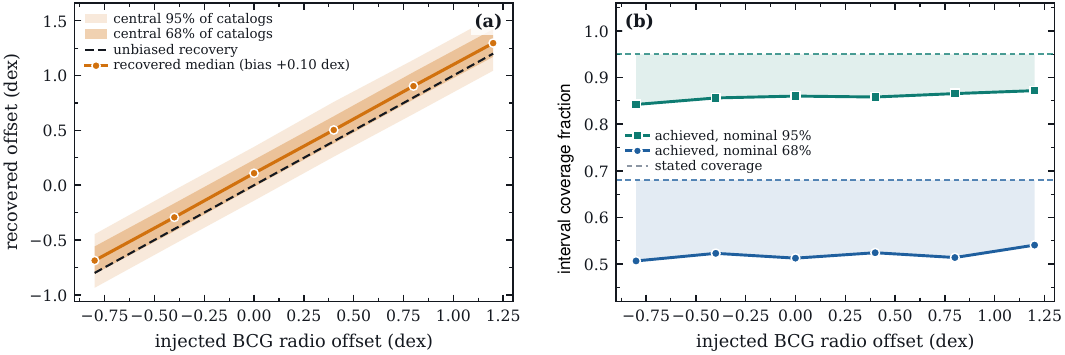}
\caption{\rev{Validation of the radio-offset measurement from the 72-BCG archive, using 3000 mock catalogs at each fixed input offset. \textit{Left:} recovered versus injected BCG radio offset. The black dashed line shows unbiased recovery; the solid curve shows the median recovered value, which lies approximately $+0.10\dex$ above it. Shaded regions contain the central $68\%$ and $95\%$ of recovered offsets. \textit{Right:} fraction of mock catalogs for which the quoted $68\%$ or $95\%$ interval contains the true radio offset inserted into the simulation. The horizontal dashed lines show the ideal fractions, 0.68 and 0.95. The actual fractions are only about 0.50 and 0.85, respectively, showing that the archive intervals are too narrow.}}
\label{fig:calibration}
\end{figure*}

\rev{Figure~\ref{fig:calibration} tests the procedure used to infer a radio offset from the 72-BCG proxy-mass archive. Mock catalogs are generated with a known BCG radio offset, analysed with the same model as the real archive, and then compared with the inserted value. This experiment shows directly whether the archive analysis is biased and whether its quoted uncertainty intervals contain the true value as often as expected. The result motivates the cautious interpretation in Section~\ref{sec:archive}.} Table~\ref{tab:bcgs} lists the working mass-error distributions and the angular scale corresponding to 100\,pc that sets the Stage-I resolution requirement. These mass-error distributions carry systematic uncertainty beyond their quoted statistical widths, and we audit its effect directly rather than assuming it is negligible. Comparing statistical-only intervals against versions with $0.10$- and $0.20\dex$ systematic floors, shared zero-point shifts of $\pm0.20\dex$, and mass-scale stretches of $\pm0.15$, the resulting range of $\Rinfo$ for the present sample spans roughly $1$--$19\%$ across this grid, with the mass-matched design remaining the most informative choice throughout regardless of which mass-error assumption is adopted. The quoted probabilities that a given BCG exceeds $10^{10}\,\Msun$ are correspondingly dependent on the adopted mass-error model and are retained only as design diagnostics, not as secure individual classifications.

\begin{deluxetable*}{lccccccc}
\tablecaption{Working mass-error distributions for the 16 direct-mass BCGs. The asymmetric published intervals are approximated by two-sided Gaussians in $\log\mbh$ and include a $0.10\dex$ systematic floor. The last column gives the angular scale subtended by 100\,pc, which sets the Stage-I resolution requirement.\label{tab:bcgs}}
\tablehead{\colhead{BCG} & \colhead{$\log\mbh/\Msun$} & \colhead{$s_-$} & \colhead{$s_+$} & \colhead{$P(\mu>10)$} & \colhead{$D$ (Mpc)} & \colhead{100 pc ($''$)}}
\startdata
A2256 & 10.393 & 0.174 & 0.146 & 0.99 & 237.6 & 0.087 \\
A2107 & 10.350 & 0.122 & 0.116 & 1.00 & 172.2 & 0.120 \\
A2147 & 10.210 & 0.103 & 0.103 & 0.98 & 146.0 & 0.141 \\
A1775 & 10.179 & 0.147 & 0.132 & 0.88 & 297.2 & 0.069 \\
A292  & 10.140 & 0.176 & 0.147 & 0.77 & 257.8 & 0.080 \\
A160  & 10.121 & 0.174 & 0.147 & 0.73 & 179.2 & 0.115 \\
A1185 & 10.049 & 0.173 & 0.146 & 0.58 & 145.0 & 0.142 \\
A1749 & 10.021 & 0.172 & 0.146 & 0.51 & 224.8 & 0.092 \\
A2388 & 9.982 & 0.111 & 0.109 & 0.43 & 246.3 & 0.084 \\
A240  & 9.964 & 0.176 & 0.147 & 0.37 & 250.0 & 0.083 \\
A592  & 9.934 & 0.174 & 0.146 & 0.30 & 261.5 & 0.079 \\
A2319 & 9.903 & 0.178 & 0.148 & 0.23 & 220.3 & 0.094 \\
A399  & 9.857 & 0.173 & 0.146 & 0.15 & 282.0 & 0.073 \\
A634  & 9.748 & 0.152 & 0.135 & 0.03 & 113.0 & 0.182 \\
A1314 & 9.724 & 0.376 & 0.219 & 0.08 & 138.0 & 0.149 \\
A2506 & 9.447 & 0.239 & 0.175 & 0.00 & 106.4 & 0.194 \\
\enddata
\tablecomments{Masses from \citet{deNicola2025,deNicola2026}}
\end{deluxetable*}

\end{document}